\documentclass[twocolumn,fullpaper]{jpsj3}

\def\runauthor{Osamu \textsc{Sakai}, Hisatomo \textsc{Harima}
}

\newcommand{\Bk}{\mbox{\boldmath $k \:$}}

\newcommand{\BQ}{\mbox{\boldmath $Q \:$}}

\newcommand{\Bi}{\mbox{\boldmath $i \:$}}
\newcommand{\Br}{\mbox{\boldmath $r \:$}}

\def\Beqa{\begin{eqnarray}}
\def\Eeqa{\end{eqnarray}}

\def\VEP{\varepsilon}

\title{%
Band Calculations for Ce Compounds with AuCu$_{3}$-type Crystal Structure on the  basis of  Dynamical 
Mean Field Theory II. - CeIn$_{3}$ and CeSn$_{3}$ - 
}

\author{%
Osamu \textsc{Sakai}\thanks{E-mail address: sakai\_ym@star.ocn.ne.jp} and
$^{1}$Hisatomo \textsc{Harima}
}

\inst{%
National Institute for Materials Science,
Sengen 1-2-1, Tsukuba 305-0047, Japan \\
$^{1}$Department of Physics, Kobe University, Kobe 657-8501, Japan
}

\recdate{
October 12, 2011
}

\abst{%
Band calculations for Ce compounds
with the AuCu$_{3}$-type crystal structure were carried out on the basis of dynamical mean field 
theory (DMFT).
The results of applying the calculation to CeIn$_{3}$ and CeSn$_{3}$ are presented as the second in a series of papers. 
The Kondo temperature and crystal-field splitting are obtained, respectively, as 190~and 390~K (CeSn$_{3}$), 8~and 160~K (CeIn$_{3}$ under ambient pressure),
and 30~and 240~K (CeIn$_{3}$ at a pressure of 2.75~GPa).
Experimental results for the photoemission spectrum are reasonably well reproduced.
In CeSn$_{3}$, a Fermi surface (FS) structure similar to that obtained by a refined calculation 
based on the local density approximation (LDA) is obtained.
In CeIn$_{3}$, the topology of the FS structure is different from that obtained by the LDA calculation but seems to be consistent with the results of 
de Haas-van Alphen experiments.
Cyclotron mass of the correct magnitude is obtained in both compounds.
The experimental result for the angular correlation of the electron-positron annihilation radiation is reasonably well reproduced 
on the basis of the itinerant $4f$ picture.
A band calculation for CeIn$_{3}$ in the antiferromagnetic state was carried out, and it was shown that the occupied $4f$ state should have a very shallow level.
}

\kword{
dynamical mean field theory, band theory, de Haas van Alphen effect, 
ARPES, magnetic excitation, positron annihilation, CeIn$_{3}$, CeSn$_{3}$
}

\begin{document}
\sloppy
\maketitle

\section{Introduction}
Nonempirical band calculations for strongly correlated electron systems have been extensively developed 
on the basis of dynamical mean field theory (DMFT)~\cite{A1}.
The $4f$ electrons in Ce compounds are typical strongly correlated electrons~\cite{A1}.
Recently, a DMFT band calculation scheme for Ce compounds was developed, 
which was applied to CePd$_{3}$ and CeRh$_{3}$ in a previous paper cited as I hereafter~\cite{A2}.
In the present paper, calculations for 
CeIn$_{3}$ and CeSn$_{3}$ will be reported.
Ce compounds with the AuCu$_{3}$-type crystal structure show
a wide variety of $4f$ electronic states from the itinerant limit to the localized limit.
CePd$_{3}$ and CeRh$_{3}$ have a nonmagnetic Fermi liquid (FL) ground state at low temperatures~\cite{A1}.
CeIn$_{3}$ has an antiferromagnetic (AF) ground state with a N\'{e}el temperature of $T_{\rm N}=10$~K~\cite{A3,A4,A5,A6}.
Recently, it was found that  
$T_{\rm N}$ decreases to zero at a pressure of 2.5~GPa. 
In addition, a transition to superconductivity occurs at $T_{\rm SC}=0.2$~K~\cite{A7,A8}.
CeSn$_{3}$ is the first heavy-fermion material for which an experiment on the de Haas-van Alphen (dHvA) effect 
has demonstrated the applicability of the 
itinerant $4f$ band picture~\cite{A9,A10,A11}.
The Fermi surface (FS) structures of CeIn$_{3}$ and CeSn$_{3}$ have been extensively studied 
from both experimental and theoretical viewpoints~\cite{A12,A13}.
It will be worthwhile to check whether DMFT calculation gives the correct FS topology and the cyclotron mass obtained experimentally.

The $4f$ state splits into the $j=5/2$ ground multiplet and the $j=7/2$ excited
multiplet with a separation of about 0.3~eV owing to the spin-orbit interaction (SOI).
The multiplet shows crystal-field splitting (CFS) of approximately 100~K.
The lower $j=5/2$ multiplet splits into the ($j=5/2$)$\Gamma_{7}$ doublet 
and the ($j=5/2$)$\Gamma_{8}$ quartet in the cubic point group, 
which we hereafter call the $\Gamma_{7}$ and $\Gamma_{8}$ states, respectively.  
It is important to take account of the SOI and CFS effects in $4f$ systems.
A theory named  NCA$f^{2}$vc (noncrossing approximation including the $f^{2}$ state as a vertex correction) 
has been developed~\cite{A14,A15,A16} and combined with the linear muffin-tin orbital (LMTO) method to carry out the DMFT band 
calculation~\cite{A16,A17}.
NCA$f^{2}$vc can include CFS and the SOI effect and also the 
correct exchange process of the $f^{1} \rightarrow f^{0}, f^{2}$ virtual excitation. 
The calculation gives an accurate order of the Kondo temperature ($T_{\rm K}$).
It was shown in I that the results of DMFT band calculation show reasonable agreement with experimental results for 
the photo emission spectrum (PES), inverse PES (IPES), 
angle-resolved PES, 
and inelastic magnetic excitation in CePd$_{3}$ and CeRh$_{3}$.
CePd$_{3}$ is a material with $T_{\rm K} \sim 250$~K, and CeRh$_{3}$ is one of the compounds with the most itinerant $4f$ states.
The hybridization intensity (HI) in these compounds is very strong because the $4f$ state hybridizes with a conduction band with a high 
partial density of states: the $4d$ states of Pd or Rh.

In this paper, calculations for CeIn$_{3}$ and CeSn$_{3}$ will be reported.
The $4f$ state hybridizes with $5p$ states of In or Sn in these compounds. 
Calculated results generally show reasonable agreement with experimental results for the PES~\cite{A18,A19,A20,A21,A22,A23}
and inelastic magnetic excitation by neutrons~\cite{A24,A25,A26},
although the calculation gives somewhat higher $T_{\rm K}$ for CeIn$_{3}$ than that expected from experiments, and 
gives a lower $T_{\rm K}$ for CeSn$_{3}$ when experimental results are examined in detail.

The FS structure at low temperatures is studied. 
Calculated dHvA frequencies show agreement with experimental results, and cyclotron masses have the correct order of magnitude.
In CeSn$_{3}$, the calculated FS is almost identical to that of Hasegawa {\it et al.}~\cite{A11} based on 
the local density approximation (LDA).
Their results show 
good agreement with those of experiments~\cite{A10}.
The FS topology depends on the energy levels of $4f$ states relative to the Fermi 
energy ($E_{\rm F}$). 
In the LDA calculation, however, the energy position of $4f$ levels is depending on details of calculation~\cite{A27,A28,A29,A30,A31,A32}. 
For example, several LDA band calculations fail to give the topology of the experimentally obtained FS structure~\cite{A12,A13}.
The position of the $4f$ levels relative to $E_{\rm F}$ is robust in DMFT because CFS is relatively large 
compared with the $4f$ band dispersion.
In CeIn$_{3}$ under pressure, the calculated FS is different from the result of the LDA calculation~\cite{A3,A32,A33,A34,A35,B,B0}
A closed electron pocket that is centered at the $\Gamma$ point appears, in contrast to the hole pocket in the LDA calculation~\cite{A3}.
This FS structure is, in some sense, similar to that of LaIn$_{3}$~\cite{B1,B2,B3}, although this does not imply the localized nature of the $4f$ state.
The effective Kondo temperature markedly increases from $T_{\rm K} \sim 10$~K at ambient pressure to 30~K at about $p=2.75$~GPa. 
The topology of the FS at these two pressures is similar but is not identical even when the paramagnetic ground state is assumed at ambient pressure.

The occupation number of electrons for a fixed wave vector k (k-ONE) is calculated for CeIn$_{3}$ for comparison with 
the experimental angular correlation of the electron positron annihilation radiation (ACAR), which has been considered to indicate a localized 
$4f$ state~\cite{A36,A37}.
Observed results can also be reproduced by the $4f$ band picture used in the present calculation.

To examine the band structure in the AF state of CeIn$_{3}$, a calculation based on a LDA+U-like model is carried out.
It is shown that a shallow level for the occupied $4f$ state is needed to reproduce the dHvA frequencies.

In \S 2, we briefly give the formulation on the basis of the LMTO method.
Results of the application to CeIn$_{3}$ under pressure are shown in \S 3, 
results for CeIn$_{3}$ at ambient pressure are given in \S 4, and 
a calculation for the AF state of CeIn$_{3}$ is presented in \S 5.
The application to CeSn$_{3}$ is shown in \S 6.
A summary is given in \S 7.

\section{Formulation}

The method of calculation is described briefly because  
its details have been given in previous papers~\cite{A2}. 
 We consider the excitation spectrum of the following Hamiltonian: 
\Beqa
{\cal H}={\cal H}_{\rm LDA} 
+
\frac{U}{2}\sum_{\Bi}
(\sum_{\Gamma,\gamma}
c^{+}_{\phi^{\rm a}\Bi\Gamma\gamma}
     c_{\phi^{\rm a}\Bi\Gamma\gamma}
 -n^{\rm LDA*}_{\Bi f})^{2}.
\label{eq.Hamiltonian}
\Eeqa
Here, 
$U$ is the Coulomb constant\cite{E1} and 
$ c_{\phi^{\rm a}\Bi\Gamma\gamma}$ is the annihilation 
operator for the atomic localized state 
$
\phi^{\rm a}_{\Bi\Gamma\gamma}(\Br)
$ 
at site  $\Bi$ with  the 
$\gamma$ orbital of the $\Gamma$-irreducible representation. 
The quantity $n^{\rm LDA*}_{\Bi f}$  
is determined using the 
occupation number of the atomic $4f$ electron per Ce ion in 
the LDA calculation.
We assume that the local Coulomb interaction acts only on the orbital 
$\phi^{\rm a}_{\Bi\Gamma\gamma}$, 
and the localized $4f$ state
$\phi^{\rm a}$ is approximated 
by the band center orbital 
$\phi(-)$.

The excitation spectrum is expressed by introducing the self-energy
terms,
\Beqa
{\cal H}_{\rm DMFT}={\cal H}_{\rm LDA} 
+ \sum_{\Bi,(\Gamma,\gamma)}
(\Sigma_{\Gamma}(\VEP+{\rm i}\delta)
 +\VEP^{\rm a}_{\Gamma}-\VEP_{\Gamma}^{\rm LDA})
 |\phi^{\rm a}_{\Bi\Gamma\gamma} >
     < \phi^{\rm a}_{\Bi\Gamma\gamma} |,
\label{eq.DMFT-Hamiltonian}
\Eeqa
where $\VEP^{\rm a}_{\Gamma}$ is the single-electron energy level of 
the $4f$ state and 
$\VEP_{\Gamma}^{\rm LDA}$ is the energy level 
in the  LDA calculation.
The self-energy $\Sigma_{\Gamma}(\VEP+{\rm i}\delta)$ is calculated by solving the 
auxiliary impurity problem with the use of  
NCA$f^{2}$vc; its outline is described in the Appendix of ref.~\citen{A16}.
This method  gives an accurate order of the Kondo temperature.
The splitting of the self-energy due to the SOI and CFS effects is considered. 

The matrix equation for the Greenian is written for a given wave number vector $\Bk$ as    
\Beqa
[ zI-D_{\rm LDA}(\Bk)-\Sigma(z)]G(z;\Bk) = I,
\label{eq.G-eq}
\Eeqa
where $I$ is the unit matrix and 
$D_{\rm LDA}(\Bk)$ is the diagonal matrix of the eigenenergies of  
${\cal H}_{\rm LDA}$ with $\Bk$.
The matrix elements of $\Sigma(z)$ are obtained by calculating the self-energy operator term of eq.~(\ref{eq.DMFT-Hamiltonian}) 
in the manifold of the eigenvectors of ${{\cal H}}_{\rm LDA}$.

The density of states (DOS) on the atomic $4f$ state is given by \\
$
 \rho^{({\rm band})}_{\Gamma}(\VEP;\Bk)=-\frac{1}{\pi}\Im{\rm tr}
[\hat{O}_{\Gamma}G(\VEP+{\rm i}\delta;\Bk)]
\label{eq.rho-band}
$, 
where the projection operator is defined as
$ 
\hat{O}_{\Gamma}=\sum_{\Bi\gamma}
|\phi^{\rm a}_{\Bi(\Gamma\gamma)}><\phi^{\rm a}_{\Bi(\Gamma\gamma)}|
$.
The local DOS in the DMFT band calculation is obtained by summing $\rho^{({\rm band})}_{\Gamma}(\VEP;\Bk)$ over 
$\Bk$ in the Brillouin zone (BZ):
$\rho^{({\rm band})}_{\Gamma}(\VEP)=\frac{1}{N}\sum_{\Bk}\rho^{({\rm band})}_{\Gamma}(\VEP;\Bk)$. 
Here, $N$ is the total number of unit cells.

The self-consistent calculation in the DMFT is as follows.  
First, we calculate the self-consistent LDA band by the LMTO method; 
potential parameters are fixed to those in the LDA calculation except for the $f$ levels.  
(I) We calculate the atomic $4f$ DOS $\rho^{({\rm imp.})}_{\Gamma}(\VEP)$ ($4f$~DOS) for the auxiliary impurity 
Anderson model by the NCA$f^{2}$vc method with a trial energy dependence of the HI 
and $4f$ levels,  
then calculate the local self-energy.
(II) The DMFT band calculation is carried out using the self-energy term, 
and the local $4f$~DOS in the DMFT band is calculated.
The calculation is iterated so that the $4f$~DOS of the local auxiliary 
impurity model and the DMFT band satisfy the self-consistent conditions. 

The $4f$ level 
is adjusted 
in the DMFT self-consistent iterations under the condition that the  $4f$ occupation number has 
a given target value, $n_{f}({\rm rsl.target})$. 
The temperature dependence of the Fermi energy, $E_{\rm F}$, is neglected by fixing it at a value determined at a low temperature.
It is estimated using the occupation number obtained by the renormalized band (RNB) calculation, in which the self-energy 
is approximated by an expansion form up to the linear term in the energy variable at $E_{\rm F}$ 
(see Appendix A of I for the calculation of the total occupation number).
The target $4f$ electron number $n_{f}({\rm rsl.target})$ is imposed on the occupation number
calculated directly using the resolvents to stabilize the self-consistent iterations.

\section{CeIn$_{3}$ under Pressure
}

\subsection{Density of states}


\begin{figure}[!htb]
\begin{center}
\includegraphics[width=10cm]{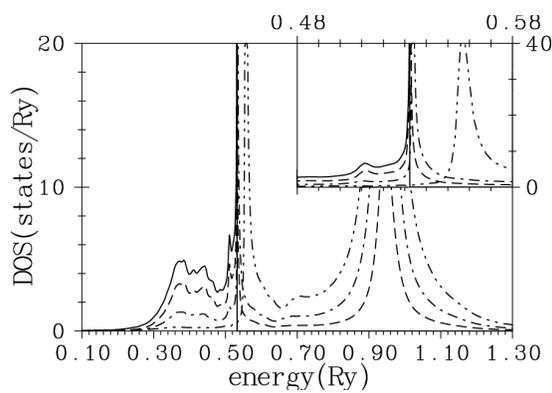}
\end{center}
\caption{
$4f$~DOS of CeIn$_{3}$ with $a=8.7326$ a.u. at $T=2.34$~K.
The solid line shows the total $4f$ PES. 
The dashed line is the DOS of the ($j=5/2$)$\Gamma_{7}$ component, 
the dot-dash line is the DOS of the ($j=5/2$)$\Gamma_{8}$ component, and 
the two-dots-dash line is the DOS of the $j=7/2$ component.
The Fermi energy $E_{\rm F}=0.53231$~Ry is indicated by the vertical dot-dash line.
The inset shows spectra in the vicinity of $E_{\rm F}$.
}
\label{fig:CeIn3-2.75GP-flsp}
\end{figure}


\begin{figure}[!htb]
\begin{center}
\includegraphics[width=10cm]{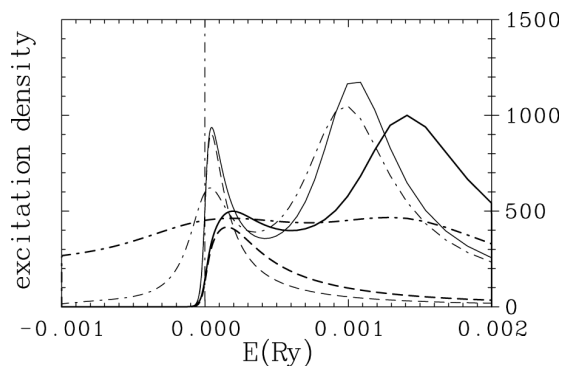}
\end{center}
\caption{
k-integrated magnetic excitation spectrum of CeIn$_{3}$. 
Solid lines show the spectrum at $T=2.34$~K and dashed lines 
show the spectrum in a hypothetical case where matrix elements of the magnetic moment are restricted 
within the intra-$\Gamma_{7}$ manifold of space.
Bold (thin) solid and dashed lines are calculated using the lattice constant 
$a=8.7326$ a.u. ($a=8.859$ a.u.).
The bold (thin) dot-dash line 
is the spectrum at $T=300$~K ($T=37.5$~K) in the case of $a=8.7326$~a.u.($a=8.859$~a.u.).
}
\label{fig:CeIn3-2.75GP-mag}
\end{figure}


\begin{table}[t]
\caption{
Various quantities obtained in the DMFT calculation for CeIn$_{3}$ with $a=8.7326$ a.u. at $T=2.34$~K.
$n_{\Gamma}^{({\rm imp.})}$ is the occupation number
in the auxiliary impurity problem of the effective HI of the DMFT calculation, 
$\VEP_{\Gamma}$ is the energy level in Ry.
$\rho_{\Gamma}(E_{\rm F})$ is the partial DOS at $E_{\rm F}$ in Ry$^{-1}$.
$\bar{Z}_{\Gamma}^{-1}=
 1-\partial\Re\Sigma_{\Gamma}(\varepsilon)/\partial\varepsilon|_{E_{\rm F}}$
is the mass renormalization factor of the $4f$ band,
$\bar{\VEP}_{\Gamma}$ is the effective energy of the renormalized band in Ry,
and $\bar{\Gamma}_{\Gamma}$ is the imaginary part of the self-energy at $E_{\rm F}$ in Ry.
($\bar{\Gamma}_{\Gamma8}$ is a very small positive value.)  
The effective energy levels are measured from the Fermi energy $E_{\rm F}=0.53231$~Ry. 
The spin-orbit interaction constant is  
$\zeta_{4f}=7.073 \times 10^{-3} $ Ry.
The $4f$ level in the band calculation is $\VEP^{\rm band}_{4f}=0.55550$ Ry.
The electrostatic CFS is 
set to be zero: $\VEP^{a}_{\Gamma 7}=\VEP^{a}_{\Gamma 8}$. 
The target $4f$ occupation number is $n_{f}({\rm rsl.target})=0.955$ and   
the resultant occupation number calculated using the resolvent is 0.955.
The $4f$ occupation number calculated by integrating the spectrum is $n_{f}({\rm intg.})=0.967$ and 
the obtained total band electron number is $N({\rm total} ; {\rm RNB})=12.999$. 
$E_{\rm inel}$ is the characteristic energy of the quasi-elastic excitation  
and $E_{\rm CFS}$ is the CFS excitation energy estimated from the magnetic excitation spectrum.
The occupation numbers of the $4f$ state in the LDA are $n_{f}({\rm Ce, LDA})=1.061$ and $n_{f}({\rm La, LDA})=0.107$.
The Coulomb constant $U$ is set to be 0.51~Ry (6.9~eV).
}
\label{tab:CeIn3-2.75GP}
\begin{halftabular}{@{\hspace{\tabcolsep}\extracolsep{\fill}}cccc} \hline
&$\Gamma_{7}$ & $\Gamma_{8}$ & $j=7/2$ 
\\ \hline
$n_{\Gamma}^{({\rm imp.})}$ & 
 0.629  & 0.263   & 0.075
                              \\
$\VEP_{\Gamma}$(Ry)                &
-0.10247  &-0.10797   &-0.15041
                               \\ 
$\rho_{\Gamma}(E_{\rm F})({\rm Ry}^{-1})$  &
50.9  & 9.8  & 0.6
                               \\ 
$\bar{Z}^{-1}_{\Gamma}$        & 
27.5     & 26.6 & 4.4 
                               \\
$\bar{\VEP}_{\Gamma}$(Ry)            &
0.0135  &0.0407 & 0.2161
                               \\  
$\bar{\Gamma}_{\Gamma}$(Ry)            &
1.99 $\times 10^{-3}$  & 0.00$\times 10^{-3}$ & 0.16 $\times 10^{-3}$
                               \\ 
\multicolumn{3}{l}
 {$E_{\rm inel}=2.7$ meV, \hspace{0.5cm} $E_{\rm CFS}=20$ meV}  & \\
\hline
\end{halftabular}
\end{table}


\begin{figure}[!htb]
\begin{center}
\includegraphics[width=10cm]{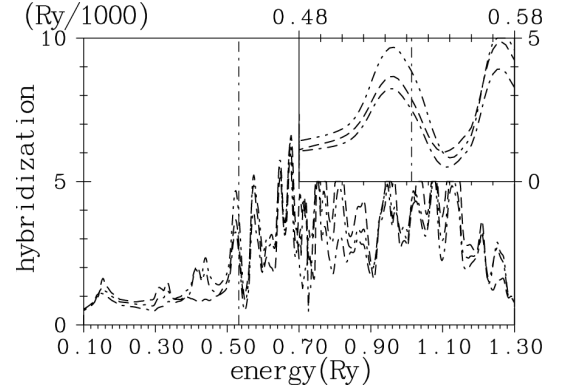}
\end{center}
\caption{
Hybridization intensity (HI) of CeIn$_{3}$ with $a=8.7326$ a.u. calculated by the LDA band method.
The dashed line is the HI of the ($j=5/2$)$\Gamma_{7}$ component, 
the dot-dash line is the HI of the ($j=5/2$)$\Gamma_{8}$ component, and 
the two-dots-dash line is the HI of the $j=7/2$ component.
$E_{\rm F}$ is indicated by the vertical dot-dash line.
The inset shows HI in the vicinity of $E_{\rm F}$.
The $4f$ levels are located at 0.71143~and 0.73620~Ry, respectively, for $j=5/2$ and $7/2$.
}
\label{fig:CeIn3-2.75GP-LDA-mix}
\end{figure}

In Fig. \ref{fig:CeIn3-2.75GP-flsp}, we show the $4f$ DOS at $T=2.34$~K
for CeIn$_{3}$ at pressure $p=2.75$~GPa with lattice constant $a=8.7326$~a.u.~\cite{A38} 
The solid line shows the total $4f$ component of the PES ($4f$~PES).
The dashed line is the DOS of the $\Gamma_{7}$ component and the dot-dash line is the 
DOS of the $\Gamma_{8}$ component.
The two-dots-dash line is the DOS of the $j=7/2$ component.
CFS of the self-energy in the excited $j=7/2$ multiplet is neglected.
The vertical dot-dash line indicates the Fermi energy $E_{\rm F}=0.53231$~Ry.
The inset shows spectra in the energy region near $E_{\rm F}$.
The $4f$~DOS has a large peak at 0.557~Ry,  about 0.025~Ry~(0.34~eV) above $E_{\rm F}$.
This peak mainly has the $j=7/2$ character.
The spin-orbit splitting on the IPES side is usually enhanced in Ce systems as noted in I.
The spectral intensity on the PES side  mostly consists of the $\Gamma_{7}$ component.
The PES has a sharp peak at $E_{\rm F}$ with a side peak at 0.021~Ry~(0.29~eV) below $E_{\rm F}$.
The latter corresponds to the SOI side band.
The PES also has a broad peak at about 0.4~Ry, which corresponds to the atomic $f^{1} \rightarrow f^{0}$ 
type excitation. 
The PES of CeIn$_{3}$ at ambient pressure has been extensively studied in experiments~\cite{A21,A22,A23}.
A comparison with these experimental results will be made in a later section.

In Fig.~\ref{fig:CeIn3-2.75GP-mag} we show the k-integrated magnetic excitation spectrum.
The total magnetic excitation spectrum at $T=2.34$~K is shown by the bold solid line, 
which has peaks at about $E=0.0002$~Ry~(32~K) and $E=0.0015$~Ry~(240~K).
The bold dashed line depicts the spectrum for a hypothetical case in which the matrix elements of the magnetic moment are nonzero only in the 
manifold of $\Gamma_{7}$, and thus it may correspond to the excitation spectrum within the $\Gamma_{7}$ manifold.
The Kondo temperature within the $\Gamma_{7}$ state is estimated to be 32~K and the CFS excitation is estimated to be 240~K.

The parameters and the calculated values are given in Table \ref{tab:CeIn3-2.75GP}.
The LMTO band parameters for states except for the $f$ component are fixed to those obtained by the LDA calculation.
$E_{\rm F}$ is fixed to the value determined by the occupied state in the RNB calculation.
The occupied $4f$ electron mainly has the $\Gamma_{7}$ character.
The occupation number of the $\Gamma_{7}$ component relative to the $\Gamma_{8}$ component, 
0.63/0.26~(2.4), is  
very large compared with 0.5 expected from the ratio of the degeneracy, 
but is smaller than the value expected from the simple model of CFS for an isolated ion with 
an excitation energy of 240~K.
The occupation of the $j=7/2$ component, 0.08, is  very small.
Usually, the electrostatic potential causes cubic CFS in $4f$ electron systems. 
This is neglected in the present calculation. 
The hybridization effect causes large CFS of 240~K. 

The target value of the occupation number on the atomic $4f$ states, $n_{f}({\rm rsl.target})$, has been tentatively set to be 
0.955 in the present calculation.
The occupation number on the atomic $4f$ state is 1.061 in the LDA band calculation for CeIn$_{3}$. 
On the other hand, the occupation number on the atomic $4f$(La) state is estimated to be 0.107 
when we carry out the LDA calculation for a compound in which Ce ions are replaced by La ions.
The difference between these two $4f$ occupation numbers is  $1.061-0.107=0.954$.
The value 0.955 is 90\% of the $4f$ occupation number (1.061) obtained by the LDA band calculation for CeIn$_{3}$.
When we perform the DMFT band calculation by setting $n_{f}({\rm rsl. target})$ to be 0.997,
94\% of the LDA value used for CePd$_{3}$~\cite{A2,A50},
$T_{\rm K}$ is estimated to be low (10~K).

In Fig. \ref{fig:CeIn3-2.75GP-LDA-mix}, we show the HI obtained from the LDA band calculation, which is used as the starting 
HI in the DMFT calculation. 
It has sharp peaks.
The partial density of states on the (In)$5p$ state has corresponding peaks, 
although they are less conspicuous.
$E_{\rm F}$ is located at a slope of a peak as indicated in the figure.
In CeIn$_{3}$, the magnitude of the Kondo temperature is very sensitive to the position of the Fermi energy, 
although the effective HI in DMFT is usually smeared out and generally reduced~\cite{A39}.
We note that an instable region appears with $\frac{\partial N_{\rm total electron}}{\partial E_{\rm F}}  < 0$ 
when $E_{\rm F}$ is reduced by 0.025 Ry.
There is a possibility that a solution with a lower $E_{\rm F}$ and an extremely low Kondo temperature appears, 
but in this paper we adopt a solution with higher $E_{\rm F}$.
Checking the relative stability of the two states and the transition between them remains as a future work.

\subsection{RNB and wave-number-vector-dependent DOS}


\begin{figure}[!htb]
\begin{center}
\includegraphics[width=10cm]{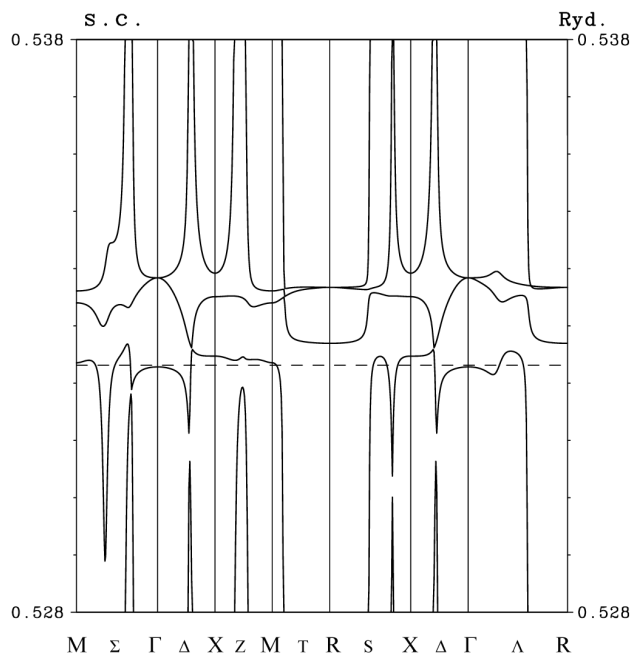}
\end{center}
\caption{
Band dispersions of the renormalized band (RNB) picture for CeIn$_{3}$ with $a=8.7326$ a.u. at $T=2.34$ K. 
The symbols under the horizontal axis denote the symmetry points and axes of the BZ of the simple cubic (s.c.) lattice.
$E_{\rm F}=0.53231$ Ry is indicated by the horizontal dashed line.
The band crossing $E_{\rm F}$ is the 7th band.
}
\label{fig:CeIn3-2.75GP-rnb}
\end{figure}


\begin{figure}[!htb]
\begin{center}
\includegraphics[width=10cm]{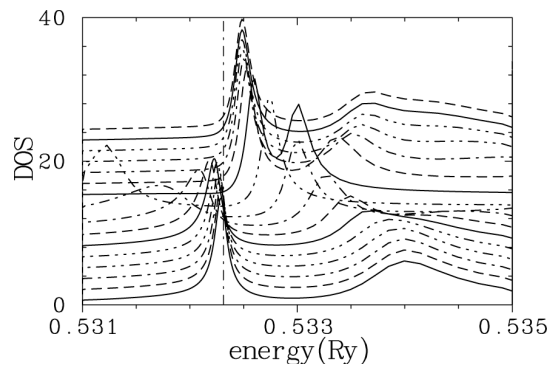}
\end{center}
\caption{
Wave number vector ($\Bk$) dependence of the DOS (k-DOS) for CeIn$_{3}$ with  $a=8.7326$~
a.u. at $T=2.34$~K. 
$\Bk$ moves from the $\Gamma$ point (bottom) to the X point (top)  along the $\Delta$ axis. 
The $4f$ DOS is shown.
$E_{\rm F}=0.53231$ Ry is indicated by the vertical dot-dashed line.
The spectra are broadened by an imaginary factor, $\gamma=0.001$~Ry, in the energy variable.
The lines are plotted in cyclic order (solid, dashed, dot-dash, two-dots-dash, and three-dots-dash lines) from the $\Gamma$ point 
but the deviations from this regularity appear  
in the peaks of the spectra because some peaks are very large.
}
\label{fig:CeIn3-2.75GP-kdos}
\end{figure}


\begin{figure}[!htb]
\begin{center}
\includegraphics[width=12cm]{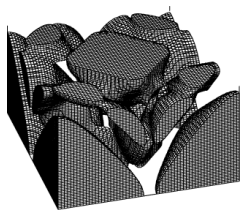}
\end{center}
\caption{
Fermi surface structure for the 7th band of CeIn$_{3}$ with $a=8.7326$~a.u. at $T=2.34$~K.
The frame is  $ -\pi/a \le (k_{x},k_{y}) \le \pi/a$ and $ -\pi/a \le k_{z} \le 0$, and the 
region occupied by the 7th band is shaded.
The center and corners of the upper plane are the $\Gamma$ point and M points, respectively.
The center and corners of the lower plane are the X point and R points, respectively.
The midpoints of the edges of the upper plane (midpoints of the neighboring M points) are also X points,
and the midpoints of the edges of the lower plane (midpoints of the neighboring R points) are also M points.
The electron pocket centered at the $\Gamma$ point is closed and the electron pocket centered at the R point 
is also closed.
}
\label{fig:CeIn3-2.75GP-FS}
\end{figure}


\begin{figure}[!htb]
\begin{center}
\includegraphics[width=8cm]{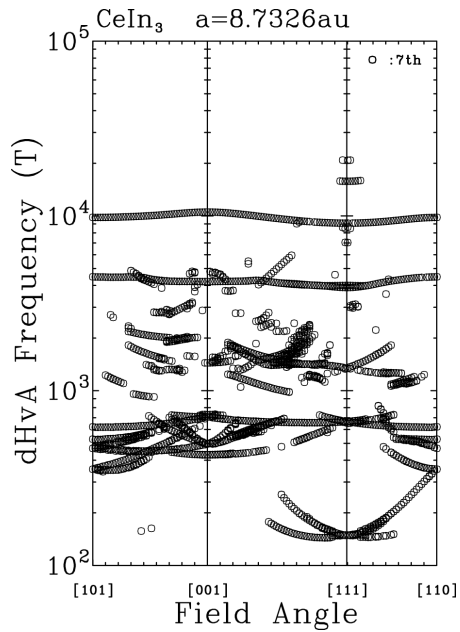}
\end{center}
\caption{
Angular dependence of the dHvA frequency of CeIn$_{3}$ with $a=8.7326$ a.u. at $T=2.34$K.
}
\label{fig:CeIn3-2.75GP-DH}
\end{figure}


\begin{table}[t]
\caption{
Calculated and experimental dHvA frequencies and cyclotron masses of CeIn$_{3}$ with $a=8.7326$~a.u.
Calculated values are estimated at $T=2.34$~K.
$F$ is the dHvA frequency in the unit of $10^{3}$~T and $m_{\rm c}$ is the cyclotron mass in the unit of electron mass $m_{0}$.
Experimental values are presented in parentheses.
$n$ in the expressions $d_{n}$ and $a_{n}$ denotes the direction of the magnetic field.
Experimental values of $a_{001}$ and $d_{001}$ are estimated from Fig. 2 of ref.~\citen{A41} at $p=2.75$~GPa.
Other experimental values are given in Table II of ref.~\citen{A3} at $p=2.7$~GPa.
}
\label{tab:CeIn3-2.75GP-DH}
\begin{halftabular}{@{\hspace{\tabcolsep}\extracolsep{\fill}}ccc} \hline
Branch  &$F$& $m_{\rm c}$ 
\\ \hline
$a_{001}$ & 
 10.5 (9.9
)  & 49.2 (50)
                              \\
$d_{001}$                &
 4.2 \ (3.7)  &10.7 \ (6)  
                               \\ 
$a_{110}$  &
9.8 \ (9.3) & 37.8 \ (52)  
                               \\ 
$d_{110}$        & 
4.5 \ (4.2)    & 28.0 \ (23) 
                               \\
$a_{111}$            &
9.1 \ (8.7)   & 48.6 \ (44) 
                               \\  
$d'_{111}$            &
4.0 \ (3.8)  & 24.5 \ (9.7)
                               \\
$d_{111}$            &
3.9 \ (3.6) & 14.4 \ (9.5) 
                               \\
\hline
\end{halftabular}
\end{table}


\begin{figure}[!htb]
\begin{center}
\includegraphics[width=10cm]{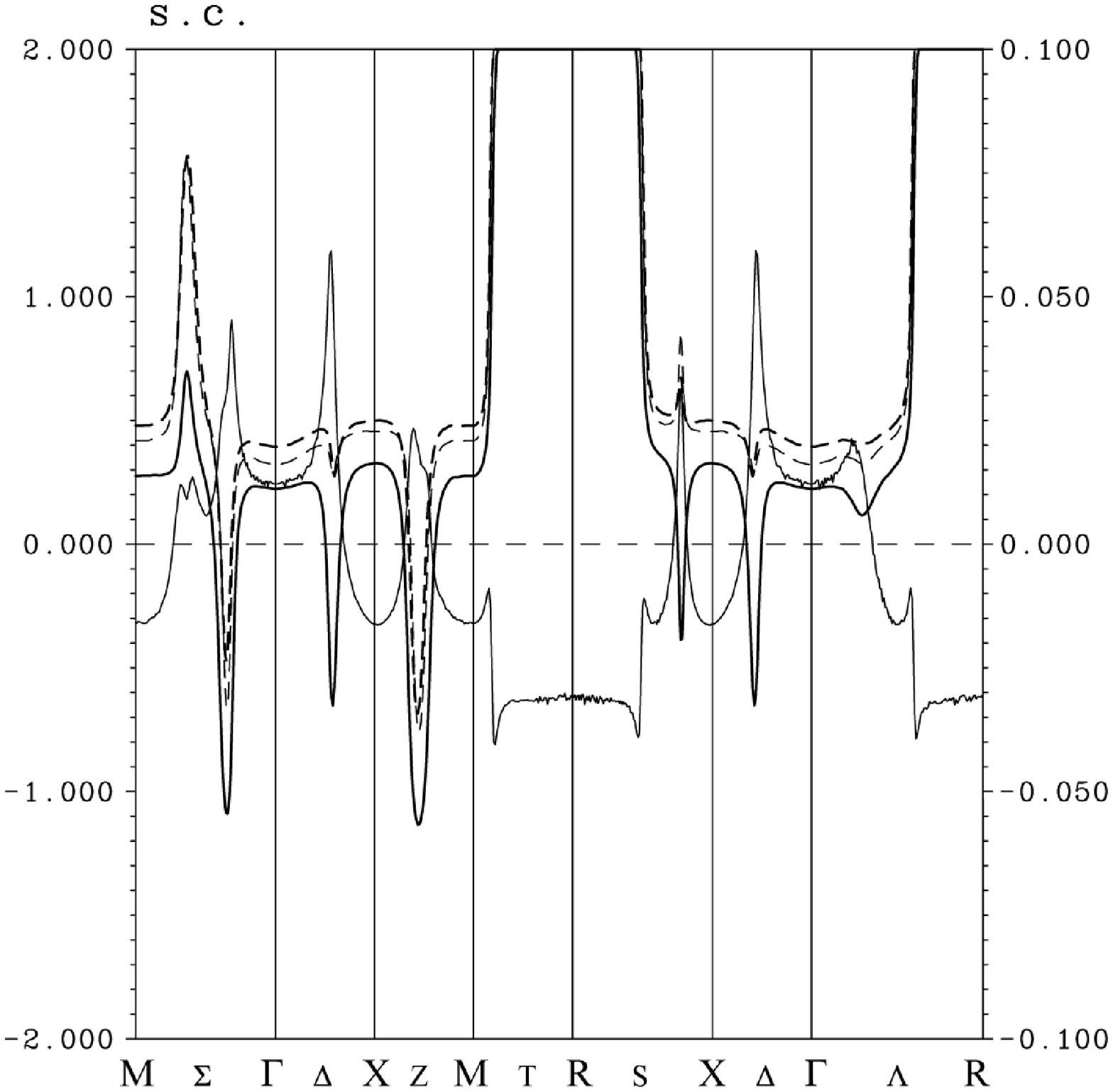}
\end{center}
\caption{
Wave number vector ($\Bk$) dependence of the occupation number of electrons (k-ONE) for CeIn$_{3}$.
Solid lines are calculated at $T=2.34$~K and the bold dashed line is calculated at $T=60$~K  in the case of $a=8.7326$~a.u.
The bold solid line is the total occupation number and the thin solid line is the occupation number of the $4f$ component.
The thin dashed line is calculated at $T=60$~K for CeIn$_{3}$ with $a=8.859$~a.u.
For the total occupation number, 13 is subtracted and the scale on the left side is used.
For the $4f$ occupation number, 1 is subtracted and the scale on the right side is used. 
In the calculation of the total k-ONE, 
an extra broadening factor of $\gamma=0.0075$~Ry is added in the k-DOS to avoid the numerical error in the 
integration of the sharp spectrum.
The k-ONE of the $4f$ component has a small random oscillation due to the numerical noise because the extra 
broadening is not added.
}
\label{fig:CeIn3-2.75GP-knos}
\end{figure}

In Fig.~\ref{fig:CeIn3-2.75GP-rnb}, we show RNB dispersions at $T=2.34$ K.
The energy shift (the real part of the self-energy at $E_{\rm F}$: $\Re\Sigma_{\Gamma}(E_{\rm F}$)) and the mass renormalization factor
($1-\partial\Re\Sigma_{\Gamma}(\varepsilon)/\partial\varepsilon|_{E_{\rm F}}$), which are 
given in Table \ref{tab:CeIn3-2.75GP}, are 
taken into account in this calculation.
The dispersions of the RNB can be understood as the hybridization bands of the $4f$ and LaIn$_{3}$-like bands.  

Narrow bands with the $j=5/2$ character appear slightly above $E_{\rm F}$.
Bands with the character of $j=7/2$ appear around an energy of 0.557~Ry, which is 
the energy of the $j=7/2$ peak in the $4f$~DOS shown in Fig. \ref{fig:CeIn3-2.75GP-flsp}, although they are not shown in 
Fig.~\ref{fig:CeIn3-2.75GP-rnb}.
The $4f$ bands located at about 0.0012~Ry above $E_{\rm F}$ mainly have the $\Gamma_{8}$ character,  
and the $4f$ bands located at $E_{\rm F}$ mainly have the  $\Gamma_{7}$ character.
The latter is the 7th band in the present calculation.
The  $\Gamma_{7}$ state sinks below $E_{\rm F}$ near the $\Gamma$ point.
This is different from the result of the LDA calculation, in which the $\Gamma_{7}$ state is located above $E_{\rm F}$ at the $\Gamma$ point\cite{A3,A32,A35}. 
We note that dispersions of the RNB are qualitatively similar to the band structure determined by the 
LDA+U calculation with the shifted $4f$ level~\cite{A40},
but the width of the $4f$ band with $j=5/2$ is about 0.005~Ry in the RNB, whereas that of the LDA+U band is about 0.02~Ry.

The dispersions of the RNB correspond well to the behavior of the $\Bk$-dependent density of states (k-DOS) as already noted in I.
For example, we show the $4f$ component of the k-DOS when $\Bk$ moves from the $\Gamma$ (bottom) to the X (top) point along the $\Delta$ line 
in Fig. \ref{fig:CeIn3-2.75GP-kdos}. 
A peak of the DOS with mainly the $4f$ character is located below $E_{\rm F}$ at the $\Gamma$ point.
Starting from the peak at the $\Gamma$ point, a ridge line runs a short distance along $E_{\rm F}$, 
and then suddenly drops on the low-energy side 
because the $4f$ band forms a hybridized band with a band of (In)$5p$ character, which has strong dispersion.
Near the X point, a ridge line with weak dispersion appears again above $E_{\rm F}$. 
Note that we have plotted the $4f$ component, not the total intensity.
If we plot the latter, it becomes an order of magnitude larger than that of the $\Gamma$ point 
for the wave number vector $\Bk \sim 0.6 \times \frac{\pi}{a}$ (the wave number vector corresponding to the 3rd solid line from the 
bottom) in the vicinity of $E_{\rm F}$ because the contribution of the (In)$5p$ component is large.
The spectral intensity is greatly reduced to about $Z \sim 1/30$ when the $4f$ component is dominant.

In Fig.~\ref{fig:CeIn3-2.75GP-FS}, we show the FS structure of the 7th band calculated from the RNB dispersion of DMFT at T=2.34K.
Two electron pockets appear, 
a smaller one centered at the $\Gamma$ point and larger one centered at the R point. 
A cagelike connected-electron sheet exists in the space between the two electron pockets.
A similar FS structure has been obtained from the LDA+U calculation with the shifted $4f$ level~\cite{A40}.

The small electron pocket at the $\Gamma$ point is closed, 
i.e., it does not touch other sheets.
On the other hand, the pocket touches the cagelike sheet in the LDA.
In addition, it has a hollow (a hole pocket) inside it in the LDA calculation 
because the $\Gamma_{7}$ state is located above $E_{\rm F}$ at the $\Gamma$ point~\cite{A3}. 
The large electron pocket at the R point is closed in both the DMFT and LDA calculations~\cite{A3}.

In Fig.~\ref{fig:CeIn3-2.75GP-DH}, we show the dHvA frequency calculated from the RNB dispersion in the DMFT calculation.
The electron pocket at the $\Gamma$ point gives signals at about $4 \times 10^{3}$~Tesla (T)
and the large electron pocket centered at the R point gives signals at about $1 \times 10^{4}$~T.
These signals correspond to the $d$ and $a$ branches in experimental results~\cite{A3,A41}.
The dHvA frequency ($F$) and cyclotron mass ($m_{\rm c}$) for magnetic fields along the principal axes are listed in Table~\ref{tab:CeIn3-2.75GP-DH}.
The calculated frequencies show good agreement with experimental results, although they are somewhat larger.
The agreement of the cyclotron masses is less good, but the calculation gives a comparable magnitude. 
Many other branches appear as seen in Fig.~\ref{fig:CeIn3-2.75GP-DH}.
For example, in the case of $H \parallel (001)$ we have 
a branch with $F=1.4 \times 10^{3}$ ($m_{\rm c}=63m_{0}$), two branches with  $F = 0.7 \times 10^{3}$ ($m_{\rm c}= 12m_{0},16m_{0}$),
a branch with $F=0.5 \times 10^{3}$ ($m_{\rm c}=7m_{0}$), and a branch with $F=0.4 \times 10^{3}$ ($m_{\rm c}=16m_{0}$).
The curvature factor (the second derivative of the cross-section area) of these branches is not so large.
However, the calculated results are less 
definite because they depend strongly on fine structures of the cagelike FS.

Recently, the momentum density of occupied electrons has been extensively studied by the ACAR.
Biasini and co-workers have concluded that the momentum density of CeIn$_{3}$ in the paramagnetic state coincides with the predictions of 
band calculations in which the $4f$ electrons are regarded as localized~\cite{A36,A37}. 
For example, let us imagine unhybridized LaIn$_{3}$-like bands by hypothetically removing the $4f$ components 
from Fig.~\ref{fig:CeIn3-2.75GP-rnb}.
When $\Bk$ moves from $\Gamma$ to X along the $\Delta$ line,
a conduction band cuts the Fermi energy from downward to upward, then another band cuts it from upward to downward.
A similar band dispersion appears around the $\Gamma$ point.
Therefore, a small electron pocket appears with a characteristic 
shell-like hole region around it in the LaIn$_{3}$-like band~\cite{A3,A12,B4}. 
A valley of momentum density appears halfway between $\Gamma$ and X points. 
Such a valley is observed in the experimental results of ACAR~\cite{A37}.
On the other hand the LDA band of CeIn$_{3}$ shows a sharp increase halfway from the $\Gamma$ point to the X point, or to the R point 
because the electron pocket at the $\Gamma$ point has the hollow of the 
hole pocket inside it~\cite{A3}.

In Fig.~\ref{fig:CeIn3-2.75GP-knos}, we show the total k-ONE on the symmetry points and axes 
obtained by the DMFT band calculation 
by a bold solid line. 
We note that k-ONE has a valley  when $\Bk$ moves from the $\Gamma$ point to the X point.
In the dispersions shown in Fig.~\ref{fig:CeIn3-2.75GP-rnb}, the $\Gamma$ point is included in the electron pocket and the  X point is included in the hole region.
Therefore, k-ONE would sharply decrease at the point where the band crosses the Fermi energy when $\Bk$ moves 
from the $\Gamma$ point to the X point along the $\Delta$ line 
if the band dispersions in Fig.~\ref{fig:CeIn3-2.75GP-rnb} were dispersions of the noninteracting bands.
The k-ONE given in Fig.~\ref{fig:CeIn3-2.75GP-knos} does not always show variations corresponding to the intersection of the RNB band with the Fermi energy.
This is partly ascribed to the large mass renormalization factor of $4f$ bands: the discontinuity of k-ONE is 
reduced to $ Z \sim 1/30$ when the crossing hybridized bands mainly have the $4f$ character. 
k-ONE shows a steep but nonsingular variation when a LaIn$_{3}$-like band cuts the Fermi energy.
Another factor may be the extra broadening introduced to avoid a numerical error in the integration of the sharp spectrum of the k-DOS.
In Fig.~\ref{fig:CeIn3-2.75GP-knos}, we also show the variation in k-ONE of the partial k-DOS of the $4f$ component by the thin solid line. 
In this calculation, only the lifetime broadening of the DMFT band state is considered, and the extra broadening factor is not 
added because the $4f$ k-DOS has a larger width.
We can recognize a relatively small but rather rapid variation of k-ONE corresponding to the Fermi function.~\cite{A42}
We may conclude that the main variation of k-ONE is determined by the dispersion of the 
hypothetical LaIn$_{3}$-like band.
The calculated result in DMFT does not appear to contradict the experimental results of ACAR.
We have also calculated k-ONE at higher temperatures and also at ambient pressures.
The total k-ONE in these cases is similar to the result shown by the bold solid line 
in Fig.~\ref{fig:CeIn3-2.75GP-knos}, except that the rapid variation 
in the $4f$ component is smeared.

We have noted that the calculated dHvA frequency is consistent with experimental results.
The dHvA frequency reflects the singular change in the momentum density, thus it can detect the FS even when the mass renormalization factor 
is very large.
An ACAR experiment with a very fine resolution at very low temperatures should be carried out.     
The calculation of ACAR including actual momentum density remains as a future work.

When we calculate the RNB dispersion at $T=150$ K, the $4f$ band shifts slightly to the high-energy side. 
This shift of the $4f$  band with increasing temperature has already been noted in I.
The $4f$ state at the $\Gamma$ point shifts to above $E_{\rm F}$. 
Therefore, 
a hole pocket appears inside the electron pocket centered at the $\Gamma$ point.
The primary structure of the FS of the RNB at $T=150$~K is similar to that of 
the LDA band of CeIn$_{3}$, not to that of LaIn$_{3}$,
except for some differences in the fine topology.
Note that the RNB picture has only limited meaning at high temperatures because the imaginary part 
is large. 
However, the trace of broad peaks in the k-DOS shows a shift corresponding to that of the RNB dispersion
as already shown for CePd$_{3}$ in I.
The calculated k-ONE does not show any indication of the hole pocket at the $\Gamma$ point even at high temperatures, as already noted.

\section{CeIn$_{3}$ at Ambient Pressure
}



\begin{figure}[!htb]
\begin{center}
\includegraphics[width=10cm]{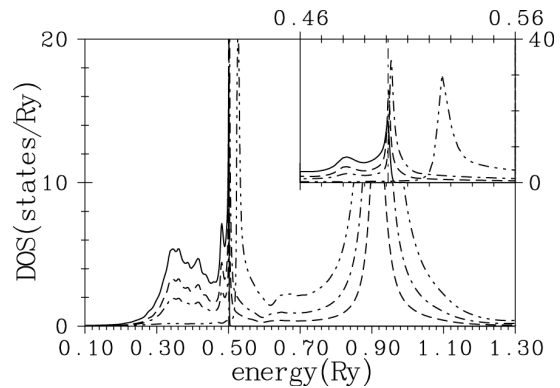}
\end{center}
\caption{
$4f$~DOS of CeIn$_{3}$ with $a=8.859$ a.u. at $T=100$~K.
For the definition of lines, see the caption of Fig.~\ref{fig:CeIn3-2.75GP-flsp}.
The Fermi energy $E_{\rm F}=0.50088$~Ry is indicated by the vertical dot-dash line.
}
\label{fig:CeIn3-0GP-flsp}
\end{figure}

\begin{table}[t]
\caption{
Various quantities obtained in the DMFT calculation for CeIn$_{3}$ with $a=8.859$~a.u. at $T=2.34$~K.
For the definition of notation, see the caption of Table \ref{tab:CeIn3-2.75GP}.
The Fermi energy is $E_{\rm F}=0.50088$~Ry and $\Delta E_{8}=0$~K.
The spin-orbit interaction constant is  
$\zeta_{4f}=7.057 \times 10^{-3}$~Ry.
The $4f$ level in the band calculation is $\VEP^{\rm band}_{4f}=0.52212$~Ry.
The target $4f$ occupation number is $n_{f}({\rm rsl.target})=0.969$ 
and the resultant occupation number calculated using the resolvent is 0.969.  
The $4f$ electron number calculated by integrating the spectrum is
$n_{f}(\rm intg.)=0.977$ 
and the calculated total band electron number is $N({\rm total} ; {\rm RNB})=13.000$.
$n_{f}({\rm Ce,LDA})=1.076$ and $n_{f}({\rm La,LDA})=0.104$.
The Coulomb constant $U$ is set to be 0.51~Ry~(6.9~eV).
$\bar{\Gamma}_{\Gamma8}$ is a very small positive value.
}
\label{tab:CeIn3-0GP}
\begin{halftabular}{@{\hspace{\tabcolsep}\extracolsep{\fill}}cccc} \hline
&$\Gamma_{7}$ & $\Gamma_{8}$ & $j=7/2$ 
\\ \hline
$n_{\Gamma}^{({\rm imp.})}$ & 
 0.721  & 0.199   & 0.057
                              \\
$\VEP_{\Gamma}$(Ry)                &
-0.10418  &-0.11058   &-0.15395
                               \\ 
$\rho_{\Gamma}(E_{\rm F})({\rm Ry}^{-1})$  &
44.7  & 6.9  & 0.4
                               \\ 
$\bar{Z}^{-1}_{\Gamma}$        & 
41.7     & 39.8 & 4.1 
                               \\
$\bar{\VEP}_{\Gamma}$(Ry)            &
0.0101  &0.0491 & 0.2489
                               \\  
$\bar{\Gamma}_{\Gamma}$(Ry)            &
4.36 $\times 10^{-3}$  & 0.00$\times 10^{-3}$ & 0.22 $\times 10^{-3}$
                               \\ 
\multicolumn{3}{l}
 {$E_{\rm inel}=0.7$ meV, \hspace{0.5cm} $E_{\rm CFS}=14$ meV}  & \\
\hline
\end{halftabular}
\end{table}

In this section we show calculated results for CeIn$_{3}$ with the lattice constant $a=8.859$~a.u. under ambient pressure~\cite{A43}. 
CeIn$_{3}$ enters the AF state at $T_{\rm N}=10$~K at ambient pressure~\cite{A4}, 
but we tentatively assume the paramagnetic state in the present section. 
In Fig.~\ref{fig:CeIn3-0GP-flsp}, we show the $4f$ DOS calculated at $T=100$~K.
It shows peaks of the Kondo resonance and the SOI side band near the Fermi edge and a broad peak at about 0.34~Ry, whose binding energy 
is about 0.17~Ry~(2.3~eV) from $E_{\rm F}$. 
Kim and co-workers have observed the PES at the $4d$-$4f$ and $3d$-$4f$ resonance thresholds 
and carefully separated the bulk and surface components~\cite{A21,A22}.
The bulk component of the PES has a broad peak with a binding energy 2~eV from $E_{\rm F}$.
The calculated intensity of the Kondo resonance part relative to the SOI side band peak 
is larger than the experimental value.
The magnetic excitation spectrum is shown in Fig.~\ref{fig:CeIn3-2.75GP-mag} by thin lines.
CFS is estimated to be 160K~($\sim$ 0.001~Ry $\sim$ 14~meV) from the peak position of the thin solid line, 
and $T_{\rm K}$ is estimated to be 8~K~($\sim$ 0.00005~Ry $\sim$ 0.7~meV) from the peak position of the 
thin dashed line.
In an experiment on neutron scattering at $T=20$~K, a broad peak of CFS excitation is observed at about 13~meV, and a 
quasi-elastic excitation peak with a width of about 2~meV is also observed~\cite{A26}.
In the calculation at $T=37.5$~K shown by the thin dot-dash line, the width of the quasi-elastic excitation peak is about 0.0002~Ry~(2.7~meV). 
The calculated value has the correct magnitude, but $T_{\rm K}$ seems to be somewhat higher than the experimental value.

We have carried out a DMFT band calculation at $T=2.34$~K assuming the paramagnetic state, although 
CeIn$_{3}$ enters the AF state at low temperatures.
The parameters and calculated values are given in Table~\ref{tab:CeIn3-0GP}. 
The mass enhancement factor is small compared with that when $p=2.75$~GPa, although the ratio of $T_{\rm K}$ is large, 32~K/8~K~$\sim$ 4.

The overall features of the RNB band dispersion are almost identical to those when $p=2.75$~GPa, which are shown in Fig.~\ref{fig:CeIn3-2.75GP-rnb},
but the $\Gamma_{7}$ band is located relatively deep from $E_{\rm F}$ when $p=0$. 
The $\Gamma_{7}$ state is located  below $E_{\rm F}$ at the $\Gamma$ point, and a closed small electron sheet centered at the 
$\Gamma$ point also appears under ambient pressure.
One important difference is that the $\Gamma_{7}$ band is located below $E_{\rm F}$ at the M point. 
Therefore, the large electron sheets centered at the R-point touch each other at the necks at M points.
The FS structure has been presented in Fig. 5 of ref.~\citen{A49}.
dHvA signals with frequencies of approximately $ 3.8\times 10^{3}$~T and cyclotron masses $m_{\rm c} \simeq 15m_{0}$ 
appear, which correspond to orbits on the small electron pocket centered at the $\Gamma$ point. 
This will be discussed in the next section.

In Fig.~\ref{fig:CeIn3-2.75GP-knos}, we show k-ONE at $T=60$~K under ambient pressure by the  thin dashed line.
It has a valley when $\Bk$ moves from the $\Gamma$ point to the X point. 
When $\Bk$ moves from the $\Gamma$ point to the R point, it shows a shallow minimum and a steep increase. 
The calculated result for total momentum density in DMFT does not contradict the experimental results of ACAR~\cite{A36,A37}
without the exceptional assumption of the localized $4f$ state.

\section{CeIn$_{3}$ in AF State
}


\begin{figure}[!htb]
\begin{center}
\includegraphics[width=10cm]{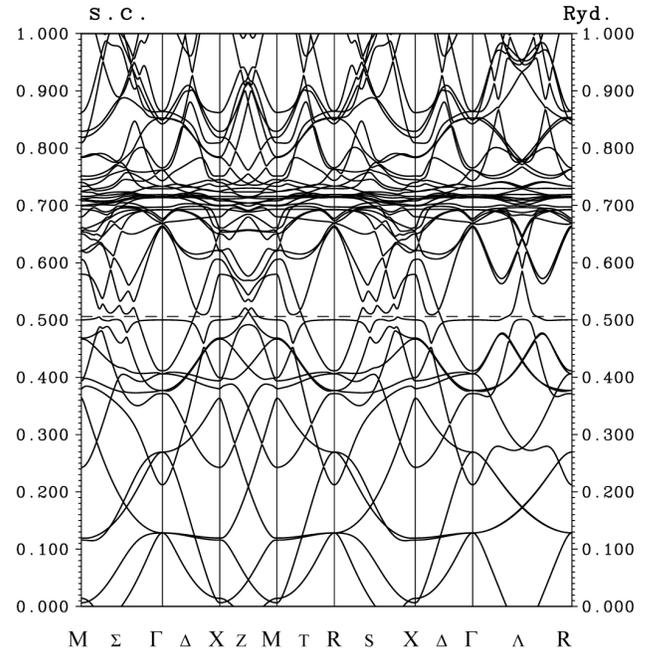}
\end{center}
\caption{
Band dispersions for the AF state of CeIn$_{3}$ with $a=8.859$~a.u. 
$E_{\rm F}=0.50610$~Ry is indicated by the horizontal dashed line.
Dispersions are depicted using the symmetry points and axes of the BZ of the s.c. lattice to illustrate the 
zone-folding effect in the AF state. 
The bands crossing $E_{\rm F}$ are the 13th and 14th bands.
}
\label{fig:CeIn3-AF-ek}
\end{figure}


\begin{figure}[!htb]
\begin{center}
\includegraphics[width=12cm]{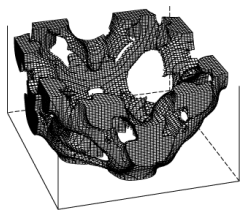}
\end{center}
\caption{
Fermi surface structure for the 13th band of the AF state of CeIn$_{3}$ with $a=8.859$~a.u.
The frame is  $ -\pi/a \le (k_{x},k_{y}) \le \pi/a$ and $ -\pi/a \le k_{z} \le 0$, and the 
hole region is shaded.
The center of the upper plane is the $\Gamma$ point and the corners of the lower plane are R points.
These two points are equivalent in the AF state. 
For the names of axes and points, see the caption of Fig.~\ref{fig:CeIn3-2.75GP-FS}.
}
\label{fig:CeIn3-AF-FS25h}
\end{figure}


\begin{figure}[!htb]
\begin{center}
\includegraphics[width=12cm]{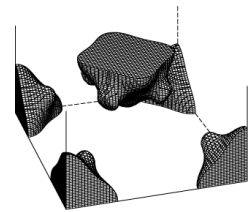}
\end{center}
\caption{
Fermi surface structure for the 14th band of the AF state of CeIn$_{3}$ with $a=8.859$ a.u.
The frame is  $ -\pi/a \le (k_{x},k_{y}) \le \pi/a$ and $ -\pi/a \le k_{z} \le 0$, and the 
occupied region is shaded.
For the names of axes and points, see the caption of 
Fig.~\ref{fig:CeIn3-AF-FS25h}.
}
\label{fig:CeIn3-AF-FS27e}
\end{figure}


\begin{figure}[!htb]
\begin{center}
\includegraphics[width=8cm]{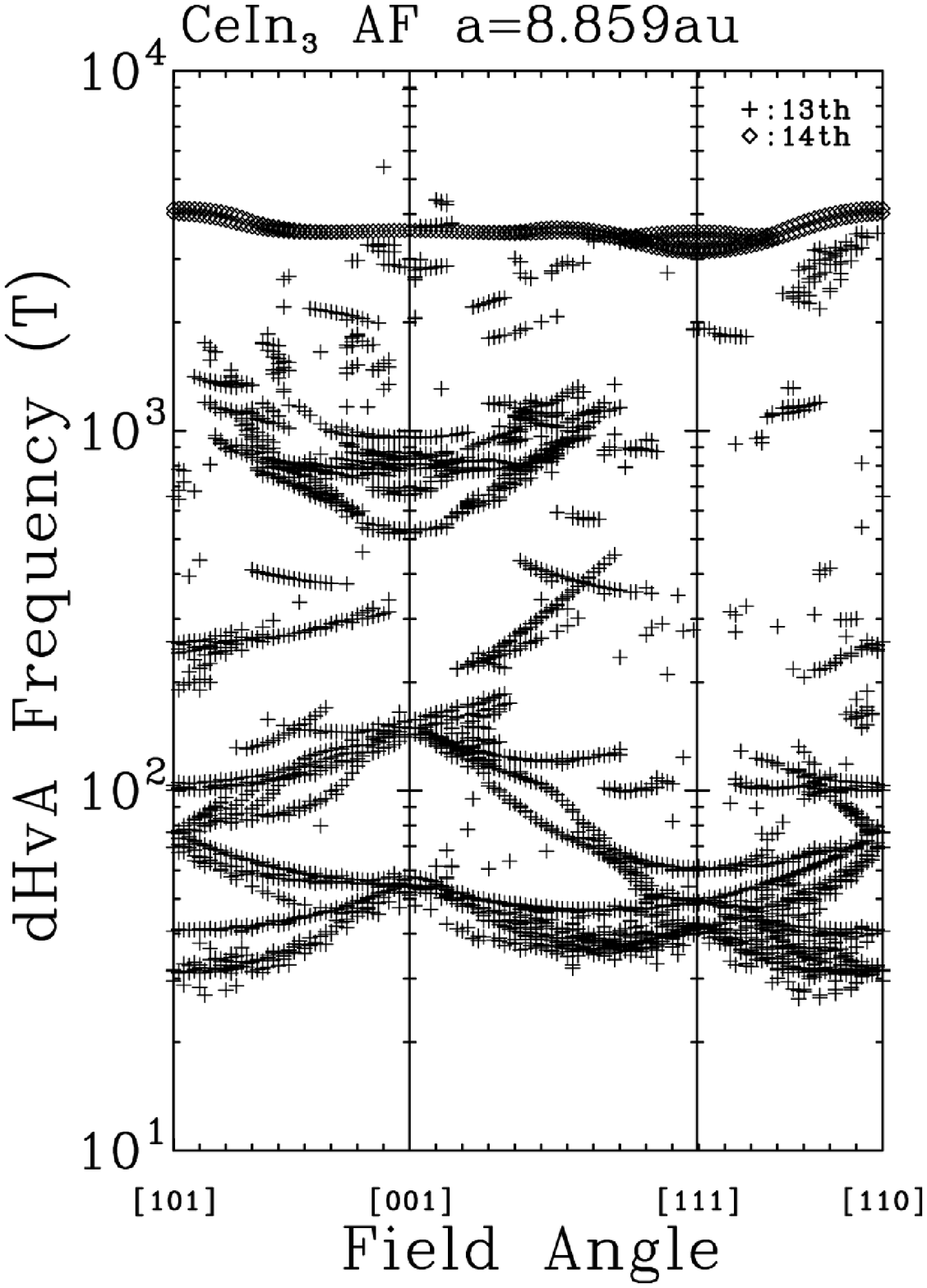}
\end{center}
\caption{
Angular dependence of the dHvA frequencies for the AF state of CeIn$_{3}$ with $a=8.859$ a.u.
}
\label{fig:CeIn3-AF-DH}
\end{figure}

At ambient pressure, 
CeIn$_{3}$ undergoes a transition to the AF state at $T_{\rm N}=10$~K. 
In this section, the band calculation assuming the AF state is carried out.
The ordering of $\Gamma_{7}$ orbitals with $\BQ=(1/2,1/2,1/2)$ is assumed. 
The orbital of the $\Gamma_{7}$ state with the polarization parallel to the (1,1,1) direction 
has low energy in a sublattice, and in another sublattice the orbital with the inverse polarization 
has low energy~\cite{A4}. 
The energy level of these occupied orbitals is chosen to be $E_{\Gamma 7}^{({\rm occ})}-E_{\rm F}=-0.0042$~Ry, where $E_{\rm F}=0.5061$~Ry.
The energy level of unoccupied states 
is set to be a higher energy of $0.2058$~Ry above $E_{\rm F}$.  
To see the characteristic features of the band structure in the AF state, a calculation similar to the LDA+U calculation is carried out 
using these energy levels but without the self-consistent condition
on U-terms~\cite{C1}. 
We note that if the energy level of whole $4f$ orbitals is set to be that of the unoccupied state, 
dispersions of the LaIn$_{3}$-like band are obtained.  
The result in the AF state is shown in Fig.~\ref{fig:CeIn3-AF-ek}.
It is depicted in the frame of the BZ of the original simple s.c. lattice, therefore, the folding 
of bands due to the AF ordering can be seen~\cite{D1}.
For example, the center of the $\Lambda$ axis is located on the zone boundary of the BZ in the AF state,
and the R point and $\Gamma$ point are equivalent.
Band dispersions are symmetric with respect to the central point on the $\Lambda$ axis~\cite{B5}. 

Note that we have chosen a very shallow energy level for the occupied $4f$ states.
When we carry out the calculation assuming deeper levels for occupied states, 
recognizable changes in band structures from those of the LaIn$_{3}$-like band do not appear 
in the energy region near $E_{\rm F}$, although new zone boundaries of the AF state are defined in a mathematical sense.
The hybridization effect in the vicinity of $E_{\rm F}$ is very small.
When we choose shallow $4f$ levels for the occupied $\Gamma_{7}$ orbitals, actual changes in band structures appear 
in the energy region near $E_{\rm F}$.
The bands showing a steep increase above $E_{\rm F}$ at the midpoint of the $\Lambda$ axis originate from the folding of a band,  
which gives the large electron pocket centered at the R point in the LaIn$_{3}$-like band. 
These steep bands with non-$4f$ character and the flat band with the $\Gamma_{7}$ character slightly below $E_{\rm F}$ show typical band 
repulsion behavior due to the $c-f$ hybridization.
On the other hand, the flat band does not show appreciable band repulsion with the 
band that forms the surface of the small electron pocket centered at the $\Gamma$ point, 
as seen on the $\Sigma$ 
and $\Delta$ axes in Fig.~\ref{fig:CeIn3-AF-ek}. 
This small electron pocket corresponds to the small electron pocket at the $\Gamma$ point in the LaIn$_{3}$-like band.
The latter has a shell-like hole sheet around it~\cite{A3}. 
The band forming the outer surface of the shell-like hole shows band repulsion due to the $c-f$ hybridization.
The hole region extends in the outward direction 
and reaches the X point.
The band dispersion of the occupied $4f$ state is very similar to the $\Gamma_{7}$ band in Fig.~\ref{fig:CeIn3-2.75GP-rnb}, 
especially for the region forming the small electron pocket centered at the $\Gamma$ point.

In Figs.~\ref{fig:CeIn3-AF-FS25h} and \ref{fig:CeIn3-AF-FS27e}, FSs are presented. 
Let us first consider the folding of the LaIn$_{3}$-like band in the BZ of the AF state.
The small electron pocket centered at the $\Gamma$ point with the shell-like hole sheet around it (see Fig. 16(a) of ref.~\citen{A3}) 
is mapped on the R point.
This hole sheet has a window (electron region) in the $\langle 111 \rangle$ direction.
The large electron pocket centered at the R point (see Fig. 16(b) of ref.~\citen{A3}) is also mapped on the $\Gamma$ point. 
The large pocket includes almost the entire shell-like hole sheet inside it.
Next we consider the $c-f$ hybridization effect  by introducing the shallow $\Gamma_{7}$ band. 
The surface of the small electron pocket is hardly changed
but the outer surface of the shell-like hole sheet moves in the outward direction away from the small electron pocket, 
i.e., the hole region becomes wide. 
In the $\langle 111 \rangle$ direction, the window of the shell-like hole 
changes to a hump of the small electron pocket.~\cite{B6}
The small electron pocket is the FS of the 14th band shown in Fig.~\ref{fig:CeIn3-AF-FS27e}. 

The surface of the large electron pocket disappears 
on the $\Delta$ axis because the corresponding 14th band is shifted up above $E_{\rm F}$ 
as seen in Fig.~\ref{fig:CeIn3-AF-ek}.
The outer surface of the shell-like hole sheet also disappears 
on the $\Delta$ axis since the 13th band is located below $E_{\rm F}$. 
The 13th band rises up near $E_{\rm F}$, but is still located slightly below $E_{\rm F}$ along the $\Sigma$ 
and S axes.
However, unoccupied regions usually exist in the 13th band in general directions.
Such a hole region can be seen on the Z axis in Fig.~\ref{fig:CeIn3-AF-ek}.
The hole sheet of the 13th band is shown in Fig.~\ref{fig:CeIn3-AF-FS27e}.
It has large  ``holes" (electron regions) along the $\langle 111 \rangle$ direction because the entire 13th band is 
located below $E_{\rm F}$ on the $\Lambda$ axis.
It also has ``holes" on the $\Delta$ and S axes.

In Fig.~\ref{fig:CeIn3-AF-DH}, we show the angle dependence of the dHvA signals, which is calculated neglecting the rotation of the 
sublattice polarization under the applied magnetic field.
The branch with $F= 4 \times 10^{3}$~T comprises the signals due to the small electron pocket (14th band), and it corresponds to the $d$ branch 
in experiments~\cite{A3,A12,A43,A44,A45,A46,A47,A48}.
The experimental dHvA frequencies (cyclotron masses) of the $d$ branch are $3.2 \times 10^{3}$~T~($2.9m_{0}$) for the field direction 
$H \parallel \langle 001 \rangle$, 
$3.6 \times 10^{3}$~T~($16m_{0}$) for $H \parallel \langle 110 \rangle$, and $2.9 \times 10^{3}$~T~($3.0 m_{0}$) 
and $3.1 \times 10^{3}$~T~($13m_{0}$) ($d'$ branch) 
for $H \parallel \langle 111 \rangle$~\cite{A3}.
The calculated values are, respectively, $3.6 \times 10^{3}$~T~($0.54m_{0}$), $4.1 \times 10^{3}$~T~($1.4m_{0}$), and $3.2 \times 10^{3}$~T~($0.52m_{0}$) and 
$3.5 \times 10^{3}$~T~($1.2m_{0}$).
The calculated frequencies show a reasonably good correspondence to the experimental values, although they are somewhat larger.
The calculated masses are smaller than the experimental values because the renormalization due to the correlation effects is not included.
We note that the small electron pocket centered at the $\Gamma$ point in the paramagnetic state also gives dHvA signals similar to that of 
the $d$ branch in the AF state.
The calculated values are, respectively, $3.7 \times 10^{3}$~T~($11.5m_{0}$), $4.0 \times 10^{3}$~T~($37.7m_{0}$), and $3.3 \times 10^{3}$~T~($12.1m_{0}$) and 
$3.5 \times 10^{3}$~T~($28.5m_{0}$).
The cyclotron masses estimated in the paramagnetic state are $3 - 5$ times larger than experimental values. 
For the field direction where the experimental values of the mass are large, 
the calculated masses are also relatively large in both the AF and  paramagnetic states.

The branch with $F \simeq 1.5 \times 10^{2}$~T for $H \parallel \langle 001 \rangle$ in Fig.~\ref{fig:CeIn3-AF-DH}
has similar angle dependence to the $s$ branch  named  by Endo {\it et al.}~\cite{A46} 
This branch runs on arms that connect sheets with a cross-section area 
having the appearance of an upside-down spade
on the $k_{x}= \pi/a$ plane in Fig.~\ref{fig:CeIn3-AF-FS25h} if the field is applied along the $x$ direction.
The centers of the orbits for $H \parallel \langle 001 \rangle $ are located at (0.10,0.12,0.38) and (0.10,0.38,0.12) and also at equivalent points in the BZ.
Many branches appear with dHvA frequencies of approximately $ 8 \times10^{2}$ and $ 5 \times 10 $~T. 
Branches with similar dHvA frequencies have been observed in experiments.~\cite{A12,A46,A47,A48} 
However, their calculated angle dependence of them does not appear to match observed ones. 
In the present calculation, we have not found  signs of hole pockets with the shape of oblate ellipsoids, which were predicted in ref.~\citen{A48}

The renormalization effect has been not included in the calculation in the present section.
A very shallow $4f$ level is needed to realize the AF effect on the band structure.
This indicates that the DMFT calculation must even be performed for the AF state in a future work.
The FS structure corresponding to the $d$ branch  should not change drastically, even in such a calculation,
because the $c-f$ hybridization has a weak effect on it.
On the other hand, the FS structure of the 14th band will change.

\section{CeSn$_{3}$
}



\begin{figure}[!htb]
\begin{center}
\includegraphics[width=8cm]{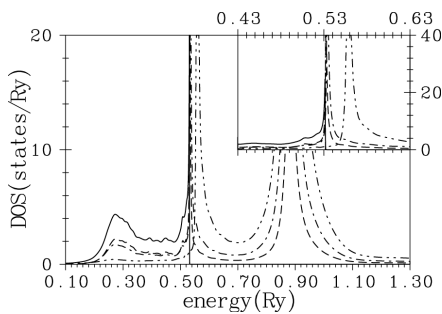}
\end{center}
\caption{
$4f$~DOS of CeSn$_{3}$ at $T=18.75$ K.
For the definition of lines, see the caption of Fig.~\ref{fig:CeIn3-2.75GP-flsp}.
$E_{\rm F}=0.53160$~Ry is indicated by the vertical dot-dash line.
}
\label{fig:CeSn3-flsp}
\end{figure}


\begin{figure}[!htb]
\begin{center}
\includegraphics[width=8cm]{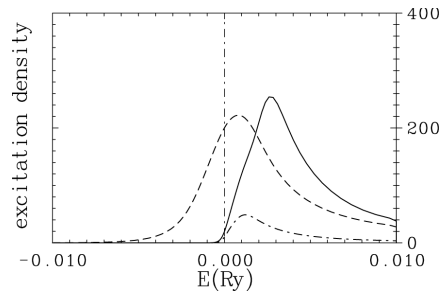}
\end{center}
\caption{
k-integrated magnetic excitation spectrum of CeSn$_{3}$. 
Solid line shows the spectrum at $T=18.75$~K and the dot-dash lines 
shows the spectrum in a hypothetical case where matrix elements of the magnetic moment are restricted 
within the intra-$\Gamma_{7}$ manifold of space.
The dashed line 
is the spectrum at $T=300$~K.
}
\label{fig:CeSn3-mag}
\end{figure}


\begin{figure}[!htb]
\begin{center}
\includegraphics[width=10cm]{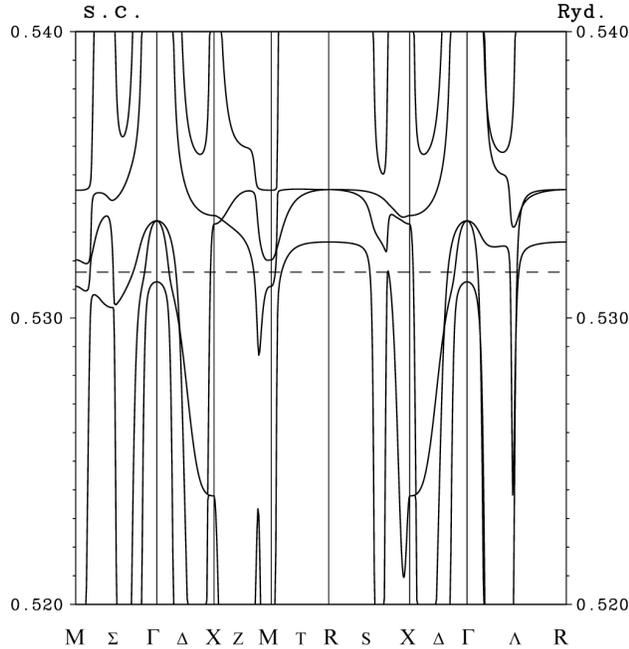}
\end{center}
\caption{
Band dispersions in the renormalized band (RNB) picture for CeSn$_{3}$ at $T=18.75$~K. 
$E_{\rm F}=0.53160$~Ry is indicated by the horizontal dashed line. 
The 4-fold (2-fold) degenerate state at about 0.5334~Ry~(0.5312~Ry) at the $\Gamma$ point is the $\Gamma_{8}$ ($\Gamma_{7}$) state.
The bands crossing $E_{\rm F}$ are the 8th and 9th bands.
}
\label{fig:CeSn3-rnb}
\end{figure}


\begin{table}[t]
\caption{
Various quantities obtained in the DMFT calculation for CeSn$_{3}$ with $a=8.921$~a.u. at $T=18.75$~K.
For the definition of notation, see the caption of Table \ref{tab:CeIn3-2.75GP}.
The Fermi energy is $E_{\rm F}=0.53160$~Ry and $\Delta E_{8}=0$~K.
The spin-orbit interaction constant is  
$\zeta_{4f}=7.008 \times 10^{-3}$ Ry.
The $4f$ level in the band calculation is $\VEP^{\rm band}_{4f}=0.54968$~Ry.
The target $4f$ occupation number is $n_{f}({\rm rsl.target})=0.98$ 
and the resultant occupation number calculated using the resolvent is 0.980.  
The $4f$ electron number calculated by integrating the spectrum is
$n_{f}(\rm intg.)=0.990$ 
and the calculated total band electron number is $N({\rm total} ; {\rm RNB})=15.995$.
$n_{f}({\rm Ce,LDA})=1.184$ and $n_{f}({\rm La,LDA})=0.167$.
The Coulomb constant $U$ is set to be 0.51~Ry~(6.9~eV).
$\bar{\Gamma}_{\Gamma8}$ is a very small positive value. 
}
\label{tab:CeSn3}
\begin{halftabular}{@{\hspace{\tabcolsep}\extracolsep{\fill}}cccc} \hline
&$\Gamma_{7}$ & $\Gamma_{8}$ & $j=7/2$ 
\\ \hline
$n_{\Gamma}^{({\rm imp.})}$ & 
 0.466  & 0.382   & 0.142
                              \\
$\VEP_{\Gamma}$(Ry)                &
-0.15825  &-0.16101   &-0.18106
                               \\ 
$\rho_{\Gamma}(E_{\rm F})$ (Ry$^{-1}$)                &
23.9  &  10.6 & 1.1
                               \\ 
 $\bar{Z}^{-1}_{\Gamma}$        & 
14.7    &12.5 & 3.3 
                               \\
$\bar{\VEP}_{\Gamma}$(Ry)            &
0.0204  &0.0422 & 0.1763
                               \\  
$\bar{\Gamma}_{\Gamma}$(Ry)            &
 $0.78 \times 10^{-3}$  & $0.0 \times 10^{-3}$ & $0.16 \times 10^{-3}$
                               \\ 
\multicolumn{3}{l}
 {$E_{\rm inel}=16$ meV, \hspace{0.5cm} $E_{\rm CFS}=34$ meV}  & \\
\hline
\end{halftabular}
\end{table}


\begin{table}[t]
\caption{
Calculated and experimental dHvA frequencies and cyclotron masses of CeSn$_{3}$.
Calculated values are estimated at $T=18.75$~K.
$F$ is the dHvA frequency in the unit of $10^{3}$~T and $m_{\rm c}$ is the cyclotron mass in the unit of electron mass $m_{0}$.
Experimental values are presented in parentheses.
The labeling of branches follows that of ref.~\citen{A11}.
The experimental values of dHvA frequency are taken from ref.~\citen{A10} and the cyclotron masses are taken from ref.~\citen{A12}. 
}
\label{tab:CeSn3-140-DH}
\begin{halftabular}{@{\hspace{\tabcolsep}\extracolsep{\fill}}ccc} \hline
branch  &$F$& $m_{\rm c}$ 
\\ \hline
$a_{001}$ & 
 8.4   & 13.6
                              \\
$a_{110}$                &
 8.4 \ (8.9)  &9.3 \ (4.2)  
                               \\ 
$a_{111}$  &
7.8 \ (8.0) & 7.5 \ (3.8)  
                               \\ 
$b_{001}$        & 
10.3 \ (9.9)    & 4.1 \ (2.7) 
                               \\
$c_{110}$            &
7.4 \ (7.4)   & 12.7 \ (6.3) 
                               \\  
$d_{111}$            &
6.4 \ (6.2)  & 11.2 \ (6.3)
                               \\
\hline
\end{halftabular}
\end{table}

In Figs.~\ref{fig:CeSn3-flsp} and \ref{fig:CeSn3-mag}, we show the $4f$ DOS and magnetic excitation spectra calculated for CeSn$_{3}$ at $T=18.75$~K,
respectively.
In the magnetic excitation spectrum, the quasi-elastic component due to the Kondo effect and the CFS component merge into a 
broad single peak at about 0.0025~Ry~(34~meV). 
However, a weak swelling due to the quasi-elastic excitation in the $\Gamma_{7}$ levels can be seen.
We estimate $T_{\rm K}$ and CFS, respectively, 
to be 190~K~(16~meV~$\sim$~0.0012~Ry) and 390~K~(34~meV~$\sim$~0.0025~Ry) from the peak positions of the dot-dash line and solid line, respectively.
A broad peak was observed at about 40 meV in an experiment on inelastic neutron scattering at T=20 K~\cite{A24,A25,A26}.
A shoulder like structure was also observed at approximately 15meV.
The calculated values show good correspondence with the experimental results. 
The calculated value of $T_{\rm K}$ is very sensitive to the lattice constant.
If we use $a=8.4515$~a.u, which was used in the calculation of Hasegawa {\it et al.}, $T_{\rm K}$ is estimated to be higher than $10^{3}$~K. 
Instead, we use $a=8.921$~a.u. which was determined by Umehara {\it et al.}~\cite{A10}, and obtained the correct magnitude of $T_{\rm K}$. 

The $4f$ DOS has a sharp peak at $E_{\rm F}$ and a small peak due to the SOI side band.
A peak at a deep energy of 0.27~Ry, whose binding energy from $E_{\rm F}$ is about 0.26~Ry~(3.5~eV) also appears. 
At first glance, the calculated spectrum is very similar to the spectrum under the $4d-4f$ resonance 
condition, but the experimental spectrum is considered to be a superposition of the surface and bulk contributions~\cite{A20,A21,A22,A23}.
The calculated spectral shape around $E_{\rm F}$ has the characteristics of the bulk component. 
On the other hand, the binding energy of the deep peak is large compared with the experimentally observed value of about 2.5eV\cite{A21}.
This is mainly due to the fact that we chose the target value of the $4f$ occupation number to be 
$n_{f}({\rm rsl.target})=0.98$.
The $4f$ occupation numbers in CeSn$_{3}$ and LaSn$_{3}$ are, respectively, 1.184 and 0.167 in the LDA calculation.
The difference between them is 1.017.
If we use $n_{f}({\rm rsl.target})=1.02$, the binding energy from $E_{\rm F}$ becomes 4.2~eV 
because we need a deep $4f$ level to increase the $4f$ occupation beyond 1.0~\cite{A50}. 
In this case, $T_{\rm K}$ and CFS are respectively estimated to be 110 and 280~K from the magnetic excitation spectrum.
When we choose $n_{f}({\rm rsl.target})=0.96$, the binding energy of the deep peak becomes 0.23~Ry~(3.1~eV).
In this case, $T_{\rm K}$ and CFS are respectively estimated to be 280 and 480~K.
However, the shoulder structure due to the SOI side band~\cite{A20,A21} becomes invisible. 
We have tentatively used $n_{f}({\rm rsl.target})=0.98$ in this paper.
The intensity ratio of the $f^{1}$ peak to the total  $f$ spectral intensity in the IPES is obtained to be 0.12, whereas 
a value of 0.18 was obtained 
experimentally~\cite{A19}.
The parameters and calculated values in the DMFT calculation for CeSn$_{3}$ are given in Table~\ref{tab:CeSn3}.
The occupation of $\Gamma_{8}$ is greater than that in the case of CeIn$_{3}$, although the CFS excitation energy is large.

In Fig.~\ref{fig:CeSn3-rnb}, the RNB is shown.
Band dispersions near the Fermi energy are similar to those of Hasegawa {\it et al.} based on the LDA calculation~\cite{A11}.
The $\Gamma_{8}$ and $\Gamma_{7}$ states are located above $E_{\rm F}$ 
at the $\Gamma$ point in their result.
The fact that the  $\Gamma_{8}$ state is located above $E_{\rm F}$ leads to the characteristic features of an electron sheet of the 9th band of CeSn$_{3}$
 (see Fig.~5 of ref.~\citen{A11}). 
The sheet is centered at the $\Gamma$ point and is essentially spherical but deeply concave in the $\langle 111 \rangle $ directions.
The eight concavities are connected with each other at the $\Gamma$ point through a small hollow when the $\Gamma_{8}$ state appears above $E_{\rm F}$.
This result is robust in the present DMFT calculation because CFS is relatively large compared with the 
dispersion of the $4f$ band. 
On the other hand, a very careful calculation is needed to obtain this result in the LDA. 
It is known that the result of Hasegawa {\it et al.} is consistent with experimental results~\cite{A10,A12,A13}.
In the present calculation, the very small hole pocket (the $p$ branch in ref.~\citen{A11}) that originates from the 7th band and is centered at the $\Gamma$ point 
does not appear because the $\Gamma_{7}$ state is located below $E_{\rm F}$, in contrast to the result of the LDA calculation.

The dHvA frequencies and cyclotron masses
are given in Table~\ref{tab:CeSn3-140-DH}.
Branch $a$ originates from the 8th band which gives a hole pocket at the R point.
The calculated dHvA frequencies show agreement with experimental values, but the cyclotron masses are about twice the experimental values.
The 8th band gives the dHvA signal of $a_{001}$ in the calculation, but it has not been observed in experiments.
The reason for this disappearance may be partly ascribed to the large cyclotron mass and large curvature factor 
as noted in ref.~\citen{A11}.
However, the mass is not particularly large compared with that of the $c_{110}$ signal, and the curvature factor 
of about 1.3 times of $c_{110}$ 
is also not particularly large. 
At present, the reason for this disappearance is not clear.

Many branches appear that originate from the 9th band.
Branches with larger dHvA frequency show correspondence with experimentally observed branches as listed in Table~\ref{tab:CeSn3-140-DH}, 
but it becomes difficult to find experimental candidates for low-frequency ones because the deviation of the frequency and angle dependence 
becomes conspicuous.

\section{Summary and Discussion}

We have studied the electronic structures of CeIn$_{3}$ and CeSn$_{3}$ 
on the basis of the DMFT calculation.
The Kondo temperatures and CFS energies are estimated to be, respectively,  8~and 160~K for CeIn$_{3}$ at ambient pressure, 
and 190~and 390~K for CeSn$_{3}$ from the k-integrated magnetic excitation spectrum. 
These values agree approximately with experimental values, but they are somewhat
higher for CeIn$_{3}$ and somewhat lower for CeSn$_{3}$. 
The magnetic excitation spectrum of CeSn$_{3}$ shows a single broad peak, but this is a 
merged peak consisting of the CFS component and the quasi-elastic component due to the Kondo effect. 
At a pressure of $p=2.75$~GPa, for which 
the volume contraction $\Delta V/V$ is about $-0.04$,
the $T_{\rm K}$ and CFS of CeIn$_{3}$ markedly increase to 32~and 240~K, respectively.
The decrease in the mass enhancement factor to about $28/42 \sim 0.7$ is, however, not particularly large.
It is noted that the Kondo temperature of CeIn$_{3}$ is very sensitive to the position of $E_{\rm F}$, 
because the HI has a very sharp peak, and $E_{\rm F}$ is located at a slope of a peak.
There is a possibility that a transition between states with different $T_{\rm K}$ appears.

In the PES of CeIn$_{3}$ at ambient pressure, 
a broad peak corresponding to the $f^{0}$ final state appears at a binding energy 2.3~eV from $E_{\rm F}$, 
consistent with the experimental value of about 2~eV.
The intensity of the Kondo resonance part relative to the SOI side band peak is somewhat larger than that 
obtained experimentally.
The calculated Kondo temperature is somewhat higher than that expected from the PES experiment.

The calculated binding energy of the $f^{0}$ peak of CeSn$_{3}$ is 3.5 eV from $E_{\rm F}$ and is 
large compared with the experimental value of 2.5~eV .
When we carry out a calculation assuming a shallower $4f$ level to reproduce the experimental binding energy, the shoulder structure of the SOI side band 
disappears.
The HI in the present calculation is larger than that expected from experiments.

In CeIn$_{3}$, the DMFT band calculation gives FS structures different from those obtained by the LDA calculation, 
whereas in CeSn$_{3}$ it gives an almost equivalent FS to that in the LDA calculation.
At the pressure $p=2.75$~GPa, CeIn$_{3}$  has two closed electron pockets, which are centered at the R point and $\Gamma$ point.
These respectively lead to the $a$ branch and $d$ branch observed in experiments.
Their calculated dHvA frequencies show correspondence with experimental results, and the cyclotron masses 
also have the correct magnitude.
In the spaces between these two electron pockets, a cagelike electron sheet appears with a complicated connected structure.
Many branches corresponding to the orbits on this FS are expected, 
although they have not yet been observed in experiments. 
The energy level of the $4f$ state gradually increases relative to $E_{\rm F}$ with increasing pressure, and thus the 
$\Gamma_{7}$ state will rise up above $E_{\rm F}$ at very high pressures.
In this case, the topology of the FS will become similar to that in the LDA calculation.

The occupation number of electrons for a fixed wave vector k (k-ONE) of CeIn$_{3}$ was calculated for comparison with the experimental result of ACAR.
When a band that mainly has the $4f$ character cuts the Fermi energy, the sharp change in k-ONE is smeared out 
at temperatures $T$ of about 60~K.
This is because the mass renormalization factor is large and the lifetime broadening is also large.
In such cases, k-ONE mainly reflects that of LaIn$_{3}$-like bands.
The calculated results do not appear to be not contradict the experimental results.
The experimental result of ACAR does not necessarily imply the localization of the $4f$ state 
in CeIn$_{3}$ at ambient pressure.

The characteristic shape of the FS of the 9th band in CeSn$_{3}$ is reproduced by the present DMFT calculation 
because the $\Gamma_{8}$ state is located above $E_{\rm F}$ at the $\Gamma$ point.
This type of band dispersion naturally occurs in DMFT since CFS is relatively large compared with the $4f$ band width.
The calculated dHvA frequencies show agreement with the results of experiments for branches with larger frequency.
The calculated cyclotron masses are about twice the experimental values for these branches. 
It is difficult to find a one-to-one correspondence of signals for lower-frequency branches.

Band dispersions for the AF state of CeIn$_{3}$ were examined on the basis of a LDA+U-like calculation.
A shallow energy level for the occupied state is needed to cause a sizable change of FS structures of the AF state 
and those of the paramagnetic state.
A small electron pocket centered at the $\Gamma$ point (and also at the R point) gives dHvA signals corresponding to the $d$ branch observed in experiments.
Calculated dHvA frequencies show agreement with those obtained experimentally, but the calculated cyclotron masses are small.
This pocket originates from the small electron pocket with a shell-like hole sheet around it in the LaIn$_{3}$-like band structure.
The small pocket centered at the $\Gamma$ point in the DMFT band of the paramagnetic state gives almost equal dHvA 
frequencies to those of the AF state.
The calculated cyclotron masses in the paramagnetic state are about $3 - 5$ times larger than the 
experimental values.
The masses are enhanced even in the AF state, but the enhancement factor is considerably reduced compared with that in the paramagnetic state.

Let us summarize the results of the DMFT band calculation for AuCu$_{3}$-type Ce compounds based on the LMTO+NCAf$^{2}$vc method.
The general features of the experimental results are reproduced by the calculations. 
The FS structures are almost identical to those obtained by the LDA calculation for CePd$_{3}$, CeRh$_{3}$, and CeSn$_{3}$. 
On the other hand, the FS structure is different from that in CeIn$_{3}$.
The result seems to be consistent with those of experiments.
Experimental findings that have been considered previously to indicate the localized $4f$ state are reproduced within 
the framework of the $4f$ band picture.
The present DMFT calculation gives a higher Kondo temperature than that obtained from the detailed analysis 
of experiments for CePd$_{3}$ and CeRh$_{3}$~\cite{B7}, 
a lower one for CeSn$_{3}$, and a considerably higher one for CeIn$_{3}$.

There is arbitrariness in the choice of the target value of the $4f$ occupation number, $n_{f}({\rm rsl.target})$. 
This induces an uncertainty in the calculated results.
Even when we examine calculations by changing this value within a range, 
it is not easy to reproduce various experimental results consistently and quantitatively.
A major origin of the discrepancy may be that the calculated HI is slightly stronger than that expected from the experimental result. 
Paying attention to these features, further applications to systems with more complex crystal structures are desirable.
Calculations of the $4f$ band state of Ce compounds in the AF state will be 
carried out in the near future.

\section*{Acknowledgments}

The author O. S. would like to thank  H. Shiba, T. Fujiwara, S. Tsuneyuki, and H. Kitazawa 
for encouragement, 
Y. Kuramoto and J. Otsuki for important  comments on the resolvent
method, and Y. Shimizu for valuable collaboration in the early stage of developing the LMTO+NCA$f^{2}$vc code. 
This work was partly supported by Grants-in-Aid for Scientific Research C (No. 21540372) and B (No. 22340099), 
an International Collaboration Research Program 
of JSPS, and  Grants-in-Aid for Research  
on Innovative Areas ``Heavy Electrons'' (Nos. 23102724 and 21102523),
and for Specially Promoted Research (No. 18002008)
from MEXT.

\vfill\eject


%


\begin{thebibliography}{9}
\bibitem{E1}
Throughout this paper we use $U=0.51$ Ry (6.9~eV), which is expected from the inverse PES of CePd$_{3}$, see ref. \citen{A2}.
\bibitem{A50}
When we used 94\% of the LDA value for CeSn$_{3}$ as in I, $n_{f}({\rm rsl.target})= 1.184 \times 0.94 =1.113$,  
we could not obtain a self-consistent solution for CeSn$_{3}$ because $n_{f}({\rm rsl. target})$ was too large.
\bibitem{B4}
On the $\Lambda$ axis, the corresponding hole region does not appear. 
The shell-like hole sheet has a ``hole" (electron region) on the $\Lambda$ axis in the LaIn$_{3}$-like band, see ref.~\citen{A3}.
\bibitem{A42}
k-ONE on the $4f$ state increases when the total k-ONE decreases because of the $c-f$ repulsion effect in the spectrum.
It shows small peaks (valleys) when the total k-ONE has valleys (peaks).
\bibitem{D1}
The translational part of the lattice structure is equivalent to fcc in the AF state. 
In a strict sense, however, the cubic symmetry is lost because of the magnetic polarization.
\bibitem{B5}
Band dispersions are also symmetric in the Z axis, i.e., the X point and M point are equivalent.
The $\Sigma$ and $\Delta$ axes are, respectively,  equivalent to the S and T axes in the AF state.
\bibitem{B6}
This can be seen in the dispersions on the $\Lambda$ line in Fig.~\ref{fig:CeIn3-AF-ek}: 
A flat band below $E_{\rm F}$ transfers to a steeply increasing band that crosses $E_{\rm F}$. 
This point is the top of the hump.
\bibitem{B7}
The target numbers of CePd$_{3}$ and CeRh$_{3}$ are chosen as 94\% of the $4f$ occupation number in the LDA calculation.
The LDA values are, respectively, 1.045 and 0.996.
\end{thebibliography}
\end{document}